\documentclass{aa}
\usepackage{graphicx}
\begin{document}

\title{New infrared star clusters in the southern Milky Way with 2MASS}

\author{ C.M. Dutra \inst{1}, E. Bica \inst{2}, J. Soares \inst{2} and B. Barbuy \inst{1} }

\offprints{C.M. Dutra - dutra@astro.iag.usp.br}

\institute{Universidade de S\~ao Paulo, Instituto de Astronomia, Geof\'\i sica e Ci\^encias Atmosf\'ericas, CP\, 3386, S\~ao Paulo 01060-970, SP, Brazil\\
\mail{}
\and
 Universidade Federal do Rio Grande do Sul, Instituto de F\'\i sica, CP\, 15051, Porto Alegre 91501-970, RS, Brazil\\
 \mail{}
 }

\date{Received --; accepted --}

\abstract{ We carried out a 2MASS J, H and K$_s$ survey of infrared star clusters   in the
Milky Way sector   230$^{\circ}$ $< \ell <$ 350$^{\circ}$. This zone was  the least studied in the literature, 
previously including only 12 infrared clusters  or stellar groups
with $|b|<$ 10$^{\circ}$,
according to the recent  catalogue by Bica et al. (2003).
We concentrated  efforts on embedded clusters, which are 
those expected  in the areas of  known radio and optical  nebulae. The present study provides 
179 new infrared clusters and stellar groups, which are interesting targets for detailed future 
infrared studies. The sample of catalogued infrared clusters and stellar groups in the Galaxy is 
now increased by 63\%.
}

\titlerunning{New infrared star clusters in the southern Milky Way with 2MASS}
\authorrunning{C. M. Dutra et al.}

\maketitle

\section{Introduction}

The development of detectors and the creation of tools to more efficiently  probe the sky in the infrared domain
are producing a rapid progress in the knowledge of the related objects (e.g. Lada \& Lada 1991, Hodapp 1994, Deharveng
et al. 1997).
In particular, near infrared imaging surveys such as the Two Micron All Sky Survey (hereafter 2MASS, 
Skrutskie et al. 1997) and the Deep NIR Southern Sky Survey (DENIS; Epchtein et al. 1997) are 
providing material for cluster identifications.  Basically these studies employed: (i) 
visual inspection of the 2MASS images (e.g. Dutra \& Bica 2000a, Hurt et al. 2000,
Dutra \& Bica 2001 and Soares \& Bica 2002), or (ii) automated searches based on stellar 
density contrast (e.g. Carpenter 2000, Ivanov et al. 2002 and Reyl\'e \& Robin 2002). 
Dutra \& Bica (2000a) scanned by eye entirely a specific area encompassing the Galactic center, whereas Dutra \& Bica (2001) 
searched for embedded clusters and stellar groups  right in the  areas of known nebulae: the method has
proven to be very efficient in the Cygnus X area and other parts of the sky.

Bica et al. (2003) gathered the infrared clusters and stellar groups  to that date into a catalogue
including 276 objects. As described therein stellar groups are less dense than clusters, but their relation
to nebulae, among other evidence, suggest them to be physical systems. In that study it became clear 
that the Southern Milky Way was the least probed sector
in the Galaxy. Recently,  two additional works report IR star cluster discoveries: Le Duigou \& Kn\"odlseder (2002) found 4 new cluster candidates in the Cygnus X area,
and Ivanov et al. (2002) with an automated search in the 2MASS Second Release Point Source Catalogue found 6 new star clusters and candidates with respect to those reported in  Bica et al.'s (2002) catalogue.
   
The present study aims at fulfilling the gap of detected IR star clusters in the Southern Milky Way by performing
a search for IR star clusters and similar objects in the direction of known nebulae, using the 
recently available 2MASS  all-sky release Atlas.
In Sect. 2 we recall the previously catalogued  clusters and
stellar groups in this Milky Way Sector. In Sect. 3 we present the newly found objects related to optical
and radio nebulae.  
In Sect. 4 we discuss some properties of the new samples. Finally, in Sect. 5 concluding remarks are given.

\section{Previously catalogued objects}

 Several star clusters  related to nebulae were previously known in the surveyed zone. For comparison
purposes with the new objects (Sect. 3) we recall their properties in the following.

Twelve infrared clusters and stellar groups   with $|b|<$ 10$^\circ$ are given in 
Bica et al. (2003), which are shown in Table 1. By columns: (1) and (2) Galactic coordinates, (3) cluster or stellar
group designation, (4) class (IRC -- infrared star cluster or IRGr -- infrared stellar Group), (5) major angular dimension, (6) distance from the Sun R, (7) multiplicity, (8) related nebulae, (9) cluster
or stellar group   major linear dimension.
  Angular dimensions correspond to  limits where object  still constrasts with respect
the background, as estimated visually on K or K$_s$ images.  
Some additional distances with respect to Bica et al. (2003)
are from studies given in Table 2 and Sect. 3. Multiplicity is a common phenomenon among infrared clusters
(Bica et al. 2002): mP means member of pair, mT triplet, m5 quintuplet etc.

\begin{table*}
\caption[]{Previously catalogued infrared clusters and stellar groups}
\begin{scriptsize}
\renewcommand{\tabcolsep}{0.9mm}
\begin{tabular}{lcccccccc}
\hline\hline
$\ell$ &   $ b$  & Cluster or Stellar Group & Class & D(')& R(kpc) & Mult. & Related nebulae & LD(pc)\\
\hline 
235.05 & -1.54 & IRAS07255-2012 Cluster  &           IRC &  1.1 &  10.2 &  mP& in distant Molecular Cloud  &                   3.3\\
235.06 & -1.55 & IRAS07255-2012 South Stellar Group& IRGr & 0.8  & 10.2 & mP & in distant Molecular Cloud   &                  2.4 \\
243.16 &  0.49 & NGC2467-East IR Cluster &           IRC &  2.9  &  4.3  & m5 & in NGC2467 Neb.,in Gum9,in Sh2-311 &            3.4 \\
265.14&   1.45 & RCW36 IR Cluster    &               IRC &  6.0  &  0.9:  & &in Gum20=RCW36=G265.1+1.5   &                   1.6 \\
267.72 & -1.09 & vdBH-RN26 IR Cluster &              IRC &  1.2 &   1.7  &  mP &in vdB-RN26=BRC59,in RCW38 &                    0.6 \\
267.92 & -1.06 &  RCW38 IR Cluster &                  IRC &  2.0  &  1.7  &  mP& in Gum22,in RCW38       &                       1.0 \\
269.18 & -1.46 & Gum25 IR Cluster &                  IRC &  3.5 &   2.6 &  & in Gum25=RCW40=G269.2-1.4    &                  2.6\\
287.81 & -0.82 &  vdBH-RN43 IR Cluster &              IRC &  1.5  &  2.5  & m10 & in vdBH-RN43,in RCW53,in $\eta$ Car Complex    &   1.1 \\
291.27 & -0.70 & NGC3576 IR Cluster&                 IRC &  2.0 &   2.4 &  mP & rel NGC3581,in G291.284-0.713,in Gum38a=RCW57a & 1.4 \\
320.15 &  0.79 & RCW87 IR Cluster    &               IRC &  2.5 &   2.6 &  mP &in RCW87,IRAS15015-5720,G320.2+0.8   &          1.9 \\
333.60 & -0.21 & G333.6-0.2 IR Cluster &             IRC  & 1.5 &  14.1 &   &in G333.6-0.2             &                     6.5 \\
336.48 &  -1.48 & RCW108 IR Cluster    &              IRC &  0.9 &   1.3  & mP&  in BRC79=G336.5-1.5,in RCW108  &               0.34 \\
\hline
\end{tabular}
\end{scriptsize}
\end{table*}

Optical open cluster catalogues can be found in Alter et al. (1970), Lyng\aa~(1987), and more recently in Dias et al.
(2002). 
Thirty seven optical open clusters appear to be embedded in optical nebulae in the present sector, as shown in Table 2.
By columns: (1) and (2) Galactic coordinates, (3) optical angular size, (4) open cluster designation, (5) multiplicity, (6) related nebulae, (7), distance $R$, (8)
age, (9) optical linear diameter, (10) class as judged from 2MASS images, (11) major angular  dimension as measured on 2MASS
images, and (12) corresponding linear dimension. 

For the sake of homogeneity we measured the optical major angular dimension on  first  (DSS) and second generation (XDSS) digitized sky survey  images using the Canadian Astronomy Data Centre extraction tool ({\rm http://cadcwww.dao.nrc.ca/cadcbin/getdss}). 
For larger angular size objects we used film copies of Schmidt  plates from the ESO red ({\rm http://archive.eso.org/wdb/astrocat/eso\_schmidt.html}) and UK blue ({\rm http://www.roe.ac.uk/ukstu/telescope.\-html}) sky surveys.   
Pairs and multiplets are very common among embedded open clusters. 
Distances and ages are
from the WEBDA database (Mermilliod 1996) -- {\rm http://obswww.unige.ch/webda/}, except those indicated in the table notes. The 
clusters are indeed young enough to be embedded or related to the nebular complexes. Cols. 10 to 12 will be discussed 
in Sect. 4.

\begin{table*}
\caption[]{Optical open clusters in nebulae}
\begin{scriptsize}
\renewcommand{\tabcolsep}{0.9mm}
\begin{tabular}{lccccccccccc}
\hline\hline
$\ell$  &    b&    D(')& Open Cluster&   Mult. &  Related Nebulae &                   R(kpc)& t(Myr)&  LD(pc)&  IR Class&  IR D(')& IR LD(pc)\\
\hline
234.78 &-0.23 &  11 &    Bochum 6  &  mP & in Sh2-309=RCW13  &                      4.0 &    10  &    13 &     -  &       - &       -\\
243.07 & 0.52 & 2.8  &   Haf19$^l$  &  m5 &    in Sh2-311   &                       5.1 &     8 &     4.2 &  IRC &       1.5 &     2.2\\
243.15 & 0.43 &  15   &  NGC2467$^l$    & - & in NGC2467 Neb.,in Gum9,in Sh2-311  &5.1$^k$ & -    &  22   &   -  &       - &       - \\
243.15 &  0.44 & 1.6    & Haf18a$^l$     & m5&  in Sh2-311,in NGC2467  &             6.0$^m$ & 8$^m$ & 2.8  & IRC$^g$  &  1.2  &    2.1\\
243.16 &  0.48 &   3   &  Haf18c$^{l,m}$ & m5& in Sh2-311,in NGC2467             &  6.0$^m$ & 8$^m$  & 5.2  &   -     &    -    &    - \\
243.17  & 0.44 & 2.1  &   Haf18b$^{l,m}$  & m5& in Sh2-311,in NGC2467 &              6.0$^k$  &-  &    3.8  & IRGr  &     1.2  &    2.1\\ 
261.45 &  0.99 &  17  &   Cr197  & -  &    in Gum15=RCW32 &                         0.8   &  13    &   4    &  -      &   -    &    -\\
262.22 & -7.81 & 370  &   Cr173   & - &    in Gum Nebula   &                      0.4$^h$ & 20$^h$  &  43  &   -   &      -  &      -\\
284.27 & -0.32 &   4  &   Westerlund 2& - & in Gum29,in RCW49 &                      1.9 &   2.5$^i$ & 2.2 &   IRC  &     2.2  &    1.2 \\
285.69 &  0.05 &   8  &   Loden 153& mP & in Gum30   &                             2.3$^i$  &  & 6.0  &  IROC  &     4   &    2.7\\
285.85 &  0.07 &   9  &   NGC3293  & - &   in Gum30   &                             2.3  &    6$^i$ & 5.4  &  -  &  -  &    -\\
286.22 & -0.17 &   7  &   IC2599$^a$  & - & in NGC3324 Neb.=Gum31=G286.195-0.163 &   2.3$^k$ & 6$^k$ & 4.7  &  IRC$^g$ &  2.8  &    1.9  \\
286.24 & -0.16 &  14  &   NGC3324  &  -  & in NGC3324 Neb.=Gum31=G286.195-0.163  &  2.3  &    6  &    9.4  &   -   &      -   &     -\\
286.76 & -1.65 &  20   &  Bochum 9  & m10&   in RCW53,in $\eta$ Car Complex  &           2.7$^k$ & -  &    16  &    -   &      -    &    -  \\               
287.02 & -0.31 &  23   &  Bochum 10  &  m10 & in RCW53,in $\eta$ Car Complex  &           2.0  &    7  &    13  &    -    &     -     &   -    \\
287.40 &-0.36  &5.5  &   Tr15   &   m10  &  in RCW53,in $\eta$ Car Complex         &    1.9  &    8   &  3.0  &   IRC$^g$  & 2.5    &  1.4         \\
287.41 &-0.57&   8  &  Tr14  &     m10 & in RCW53,in $\eta$ Car Complex    &        2.7  &   7  &  6.3 &   IRGr  &    1.9  &    1.5           \\
287.49 & -0.53 &   7 &    Cr232 &    m10 &   in RCW53,in $\eta$ Car Complex   &          3.0   &   5  &  5.5   &  IRGr  &    2.6  &    2.3  \\
287.55 & -1.02 &  29 &    Cr228  &   m10  &  in RCW53,in $\eta$ Car Complex     &        2.2   &   7  &  6.1   &   -    &     -     &   -  \\
287.59 & -0.66 &  14 &    Tr16   &   m10 &   in RCW53,in $\eta$ Car Complex    &         2.7   &   6  &   11    &  -     &    -     &   - \\
287.64 & -0.68 &   4 &    Cr234  &   m10 &   in RCW53,in $\eta$ Car Complex    &         2.7$^k$ & -  &  3.1   &   -     &    -     &   -   \\
288.03 & -0.86 &   4 &    Bochum 11 & m10 &  in RCW53,in $\eta$ Car Complex    &         2.4   &   6  &  2.8   &  IRC  &     1.7& 1.2 \\
289.50 &  0.12 &   3 &    Pis17    &  -  & in NGC3503=vdBH-RN46,in Gum34b,in RCW54& 3.5  & 4.5$^î$ & 3.1   &  IRC   &    1.5& 1.5 \\
290.33 & -2.99 &   4 &    Graham 1  &  - & includes vdBH-RN45,in RCW55=G290.3-3.0 & 3.6$^f$ & -  &  4.2   &  IRC$^g$  & 2.1 & 2.2\\
290.71 &  0.20 &   9 &    NGC3572  & - &   in Gum37=RCW54c            &             2.0 &     8  &  5.2   &   -      &   -    &    -\\
291.61 & -0.52 & 1.9 &    NGC3603  & - &   in Gum38b=RCW57b           &             3.6  & 2$^i$ &  2.0   &  IRC  &     1.8\\
294.81 & -1.64 &  11 &    IC2944   & mP  &  in Gum42=RCW62            &              1.8  &    7  &  5.8    &  -      &   -  &      -\\
294.96 & -1.71 &  11 &    IC2948  &  mP &   in Gum42=RCW62            &              1.8$^k$ & -  &  5.8    &  -       &  -    &    -\\
306.07 &  0.20 & 4.5 &    Stk16nw$^b$  & mP& in RCW75=Gum48a           &              1.9   &   5  &  2.5    &  -    &     -   &     -  \\
306.10 &  0.12 &   2 &    Stk16se$^c$ & mP &in vdBH-RN60a,in RCW75=Gum48a  &         1.9$^k$  &   &  1.1    & IRGr   &   1.4    &  0.8\\
308.67 &  0.60 & 1.5 &    BH151$^d$ & - &  in RCW79             &                   3.8$^j$ &  - &   1.7  &   IRC    &   1.2   &   1.3\\
320.51 & -1.20 & 4.5 &    Pis20  &  -   &  rel. to radio SNR MSH15-52$^e$  &        2.0  &   7   & 2.6   &  IRC    &   3.0  &    1.7\\
336.71 & -1.59 &  16 &    NGC6193 &   mP&   in RCW108=Gum53           &              1.2  &   6   & 5.6  &   IROC    &  7.0    &  2.1 \\
343.46 &  1.17 &  17 &    NGC6231  &    m4& in Gum55=RCW113           &              1.2  &   7   & 5.9   &  IROC    &  6.5    &  2.3 \\
344.39 &  1.66 &  55 &    Harvard 12 & m4 & in Gum55=RCW113          &               1.2  &   -   &  19  &    -     &    -    &    - \\
344.60 &  1.61 &   5  &   BH205$^d$   &  m4& in Gum55=RCW113            &             1.2   &   -   &  1.7    & IROC  &     5  &     1.7\\
344.94&  1.60 &  45  &   Tr24    &    m4  & in Gum55=RCW113   &                      1.1    &  8    &  14   &   -    &     -   &     -\\
\hline
\end{tabular}
\end{scriptsize}
\begin{list}{}
\item  Notes: $^a$ central part of NGC\,3324 open cluster, which is the object in WEBDA; $^b$ Stk\,16 in WEBDA; $^c$; Stk\,16 corresponding to the original coordinates (Alter et al. 1970). $^d$ van den Bergh \& Hagen (1975); $^e$ optical filaments are RCW\,89; $^f$ from Graham (1970); $^g$loose IRC; $^h$ adopted from Hipparcos; $^î$adopted from Piatti et al. (2002); $^j$ adopted from the nebula complex (Sect. 3); $^k$ adopted from other cluster in the complex; $^l$NGC\,2467 includes Haf\,19, Haf\,18a,b,c; NGC\,2467 in WEBDA corresponds to a foreground group of stars; $^m$ Haf\,18a and Haf\,18b
are respectively the north and south parts of Haf\,18 in WEBDA.  
\end{list}
\end{table*}

\section{Newly found objects}

\begin{figure} 
\resizebox{\hsize}{!}{\includegraphics{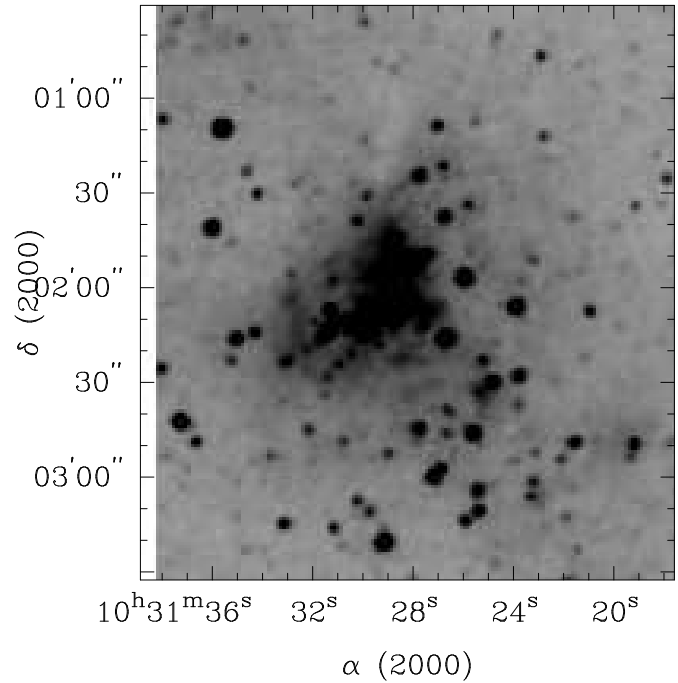}}
\caption[]{3$^{\prime}$ $\times$ 3$^{\prime}$ 2MASS K$_s$ image of prominent populous infrared cluster at J2000 $\alpha$ =  10$^h$31$^m$29$^s$ and $\delta$ = -58$^{\circ}$02$^{\prime}$01$^{\prime\prime}$  in the optical nebula Hoffleit\,18 = G285.3+0.0.}
\label{fig1}
\end{figure}

\begin{figure} 
\resizebox{\hsize}{!}{\includegraphics{h4100f2.ps}}
\caption[]{3$^{\prime}$ $\times$ 3$^{\prime}$ 2MASS K$_s$ image of prominent infrared cluster at J2000 $\alpha$ =  13$^h$11$^m$39$^s$ and $\delta$ = -62$^{\circ}$33$^{\prime}$15$^{\prime\prime}$  in the radio nebula G305.3+0.2 = AFGL4163.}
\label{fig1}
\end{figure}

\begin{figure} 
\resizebox{\hsize}{!}{\includegraphics{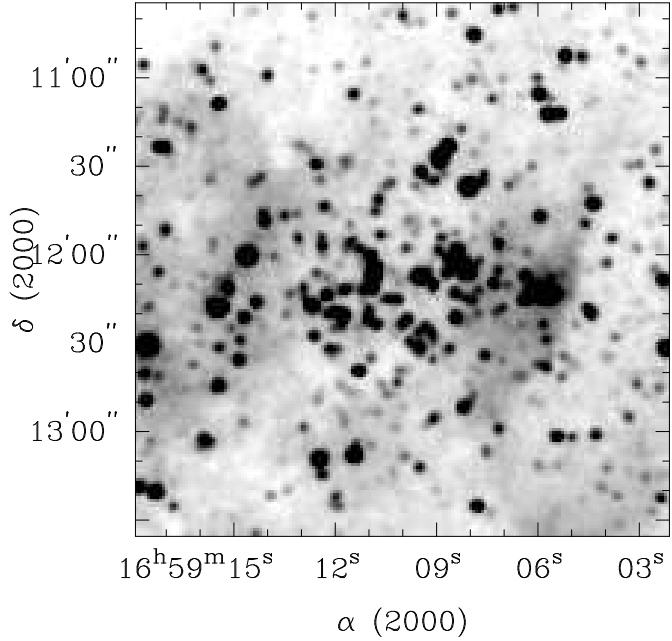}}
\caption[]{3$^{\prime}$ $\times$ 3$^{\prime}$ 2MASS K$_s$ image of loose infrared cluster at J2000 $\alpha$ =  16$^h$59$^m$10$^s$ and $\delta$ = -40$^{\circ}$12$^{\prime}$05$^{\prime\prime}$  in the  nebula G345.3+1.5.}
\label{fig1}
\end{figure}

\begin{figure} 
\resizebox{\hsize}{!}{\includegraphics{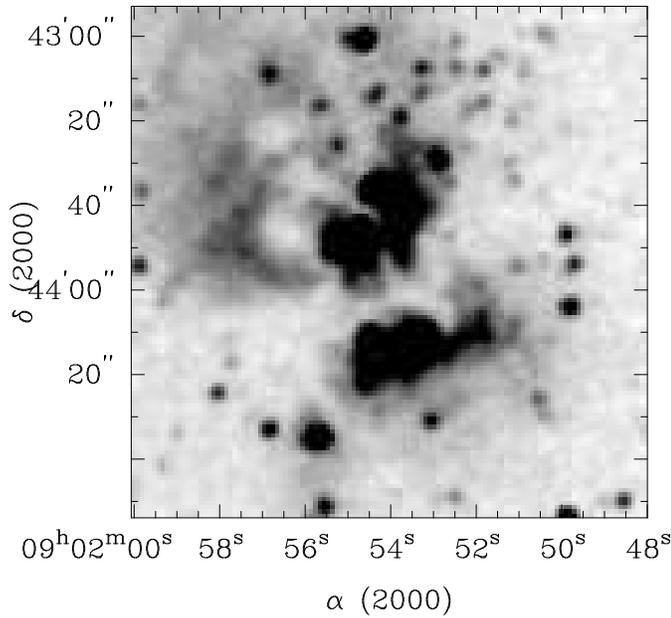}}
\caption[]{2$^{\prime}$ $\times$ 2$^{\prime}$ 2MASS K$_s$ image of deeply embedded cluster at J2000 $\alpha$ =  9$^h$01$^m$54$^s$ and $\delta$ = -47$^{\circ}$43$^{\prime}$53$^{\prime\prime}$ J2000 in the nebula Bran\,222.
The object is remarkably similar to the well-known infrared cluster in NGC\,2024.}
\label{fig1}
\end{figure}

\begin{figure} 
\resizebox{\hsize}{!}{\includegraphics{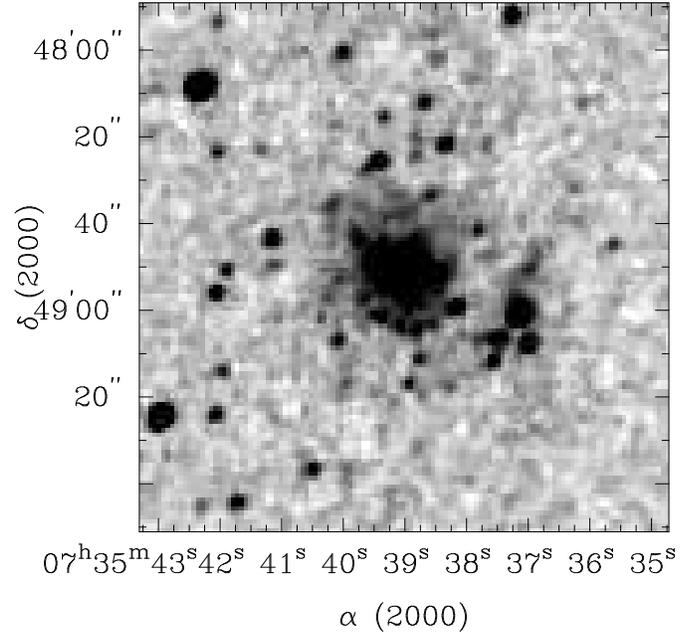}}
\caption[]{2$^{\prime}$ $\times$ 2$^{\prime}$ 2MASS K$_s$ image of infrared cluster candidate at J2000 $\alpha$ =  07$^h$35$^m$39$^s$ and $\delta$ = -18$^{\circ}$48$^{\prime}$50$^{\prime\prime}$  in the nebula Gum\,7 = Sh\,2-307 = RCW\,12.}
\label{fig1}
\end{figure}

\begin{figure} 
\resizebox{\hsize}{!}{\includegraphics{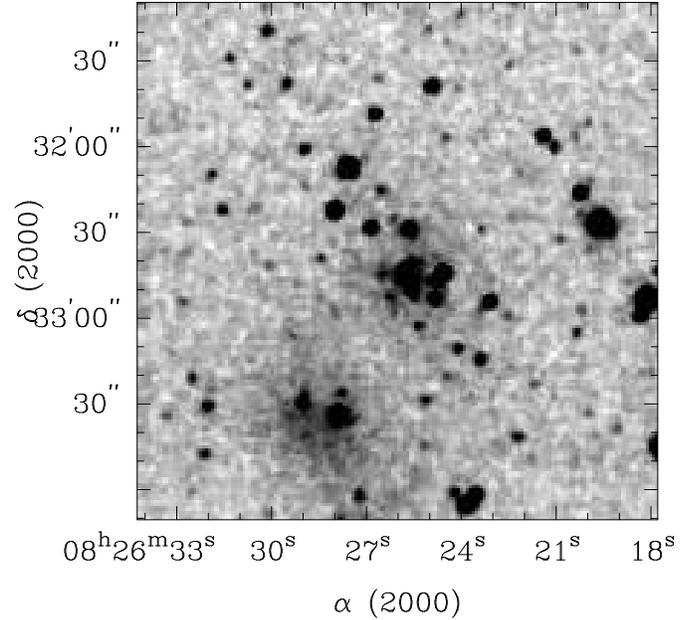}}
\caption[]{3$^{\prime}$ $\times$ 3$^{\prime}$ 2MASS K$_s$ image of infrared stellar group at J2000 $\alpha$ =  8$^h$26$^m$26$^s$ and $\delta$ = -42$^{\circ}$32$^{\prime}$40$^{\prime\prime}$  in the nebula  Bran\,149.}
\label{fig1}
\end{figure}

\begin{figure} 
\resizebox{\hsize}{!}{\includegraphics{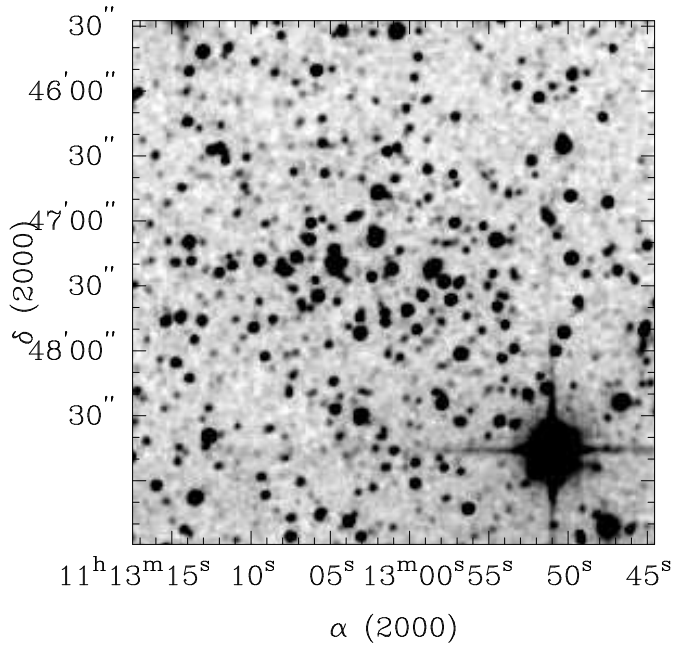}}
\caption[]{4$^{\prime}$ $\times$ 4$^{\prime}$ 2MASS K$_s$ image of infrared cluster at J2000 $\alpha$ =  11$^h$13$^m$01$^s$ and $\delta$ = -60$^{\circ}$47$^{\prime}$28$^{\prime\prime}$  with appearence of typical optical open cluster in the nebula GG\,2 = G291.2-0.2.}
\label{fig1}
\end{figure}

Since embedded clusters are expected to occur in the area of nebulae, we concentrated  search efforts
on known optical and radio nebulae, mostly HII regions but also reflection nebulae and
supernova remnants.

The search  employed  the recently available 2MASS All-Sky Release Images provided
by means of the 2MASS Survey Visualization \& Image Server facility in the web interface {\rm http://irsa.ipac.caltech.edu/}.
We extracted JHK$_s$ images with 5$^{\prime}\times$5$^{\prime}$ centred on the coordinates
of each nebula. For the nebulae with sizes larger than 5$^{\prime}\times$ 5$^{\prime}$ we took additional images of
10$^{\prime}\times$10$^{\prime}$ or 15$^{\prime}\times$15$^{\prime}$. The K$_s$ band images allow one to probe  deeper
in more absorbed regions, and the J and H band images were used mostly as control of the presence of bright stars
and as additional check for cluster resolvability.
  For the resulting IR star clusters we determined  accurate positions and dimensions from their images in FITS format using {\bf SAOIMAGE 1.27.2} developed by Doug  Mink. SAOIMAGE uses information in a 2MASS image header
to transform linear to equatorial coordinates. Centers and  angular dimensions 
are estimated visually on the 2MASS K$_s$ images. 

The optical nebula designations throughout this study are from
Ced (Cederblad 1946),
Hoffleit (Hoffleit 1953), Gum (Gum 1955), Sh2- (Sharpless 1959), RCW (Rodgers et al. 1960), GG (Georgelin \& Georgelin 1970a), GeGe (Georgelin \& Georgelin 1970b), vdBH-RN (van den Bergh \& Herbst 1975), ESO (Lauberts 1982), BFS (Blitz et al. 1982), Bran (Brand et al. 1986), BRC (Sugitani \& Ogura 1994). Some small angular size nebulae are from Wray (1966).
The  radio nebula G designations are from
various studies, namely Wilson et al. (1970), Caswell (1987) and Kuchar \& Clark (1997). MSH is from Mills et al. (1961) and references therein, and MHR from Mathewson et al. (1962). We also indicate some infrared nebulae related to sources 
in the AFGL and IRAS catalogues. 

   We merged the different catalogues and lists of nebulae into a radio/infrared and an 
optical nebula files. We cross-identified
nebulae in each  file, and then between the two files. Radio nebulae with optical counterparts were transferred to the
optical file. We also used information of optical nebulae among radio detections by Caswell \& Haynes (1987).
The resulting input lists of optical and radio nebulae contain respectively 991 and 276 objects in the present Milky Way sector, whose directions were inspected.  The whole Milky Way radio and optical nebula catalogue currently has 4454 entries after cross-identifications, and will be
provided in a forthcoming study. It follows similar  procedures as those used in the construction of the dark nebula
catalogue (Dutra \& Bica 2002) which has 5004 entries.

The results of the cluster survey will be available as Tables 3 and 4 in eletronic form at CDS (Strasbourg)
 via anonymous ftp to cdsarc.u-strasbg.fr (130.79.128.5), respectively for optical and radio nebulae. 
By columns: (1) running number; (2) and (3) Galactic coordinates, (4) and (5) J2000.0 equatorial coordinates, (6)
and (7) major and minor angular dimensions, (8) related nebulae, (9) class, (10) remarks including distance $R$ (in case of kinematical 
ambiguity the near and far distances are shown), multiplicity and linear dimension. 
The new infrared clusters, stellar groups and candidates from  the optical nebula survey amount to 123,
and from the radio nebula survey  56. The rates of detection relative to the input nebula catalogues are 12 and
20\% for optical and radio nebulae, respectively. The higher detection rate among radio nebulae is surely because
they represent more often overall complexes, while optical nebulae deal more often with structural details of closer complexes.

Object classes are infrared cluster (IRC), stellar group (IRGr), cluster candidate (IRCC),
and open cluster (IROC). IRCs are in general populous and at least partially resolved. We illustrate in Fig. 1 a prominent compact  IRC (Object 48) related to an optical nebula, and in Fig. 2 one (Object 131) related to a radio
nebula. A loose IRC (Object 114)  is shown  Fig. 3. A deeply embedded in nebular emission and/or reflection partially  resolved IRC (Object 33) is given in Fig. 4. This object
has a clear dust lane, and remarkably resembles the well-known NGC2024 cluster (e.g. Lada et al. 1991).
The infrared
cluster (Object 146) in the radio nebula G\,327.3-0.5 is also similar to that in NGC\,2024. 
IRCCs are probably clusters, but
are essentially unresolved, and require higher resolution and deeper images for definitive diagnostic (e.g. Object 8 -- 
Fig. 5). 
IRGrs are less dense than IRCs, some are rather compact but little populated. An IRGr (Object 20) is 
given in Fig. 6.  IROCs (e.g. Object 61 -- Fig. 7) have similar appearence to optical open clusters and  relatively
large angular size ($\approx$2' or more). In Table 3  there occur  46 IRC, 64 IRGr, 7 IRCC and 6 IROC objects,
while in Table 4  31 IRC, 8 IRGr, 16 IRCC and 1 IROC objects.

\begin{figure*} 
\resizebox{\hsize}{!}{\includegraphics{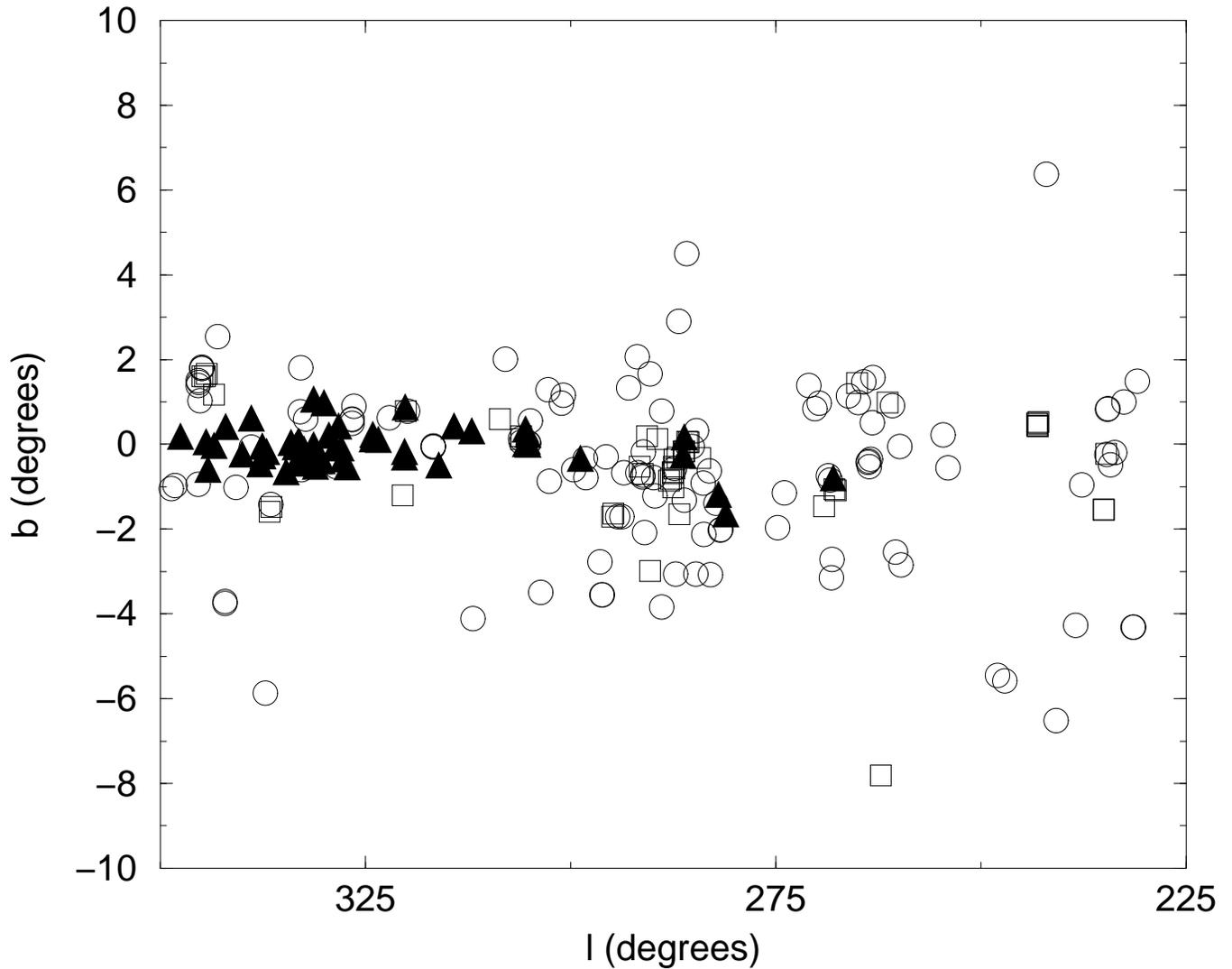}}
\caption[]{Angular distribution of infrared clusters found in the vicinity of optical nebulae (Table 3) represented by open circles,  and radio nebulae (Table 4) represented by filled triangles. Known star clusters in the region (Tables 1 and 2) are the open squares.}
\label{fig1}
\end{figure*}

   Distances are mostly based  on kinematical estimates for the nebulae (Wilson et al. 1970, Caswell \& Haynes 1987), but include as well averages
with estimates from individual stars, when available (e.g. Georgelin et al. 1973, Georgelin et al. 2000 and references therein). The near/far distance   ambiguity 
has been solved  by some of these authors ($R$), else we indicate both ($R_n$ or $R_f$).

One prominent cluster (Object 23) appears to be  related to the optical nebula Munch\,23 (Munch 1955), which is thus not  a planetary
nebula (K2-15 or PNG263.2+00.4, Acker 1992). Likewise, the nebulae ESO\,313*N10  and ESO\,128EN25 are not 
planetary nebulae. 
    
We included in Table 3 new embedded clusters and stellar groups related to optical reflection nebulae.
This object type has been discussed in Dutra \& Bica (2001). They appear to be  less massive clusters or stellar groups
where no ionizing star was formed. The objects in Soares \& Bica (2002) are of this type or close to its limit
towards ionizing stars. Detected objects in the van den Bergh \& Herbst (1975) reflection nebulae are probably of this type.

The  Gum\,38a complex (often designated NGC\,3576 H\,II region) is well-known for harbouring the prominent NGC\,3576 IR cluster (e. g. Persi et al. 1994). However, we realized that
it is not related to the  original NGC catalogue small nebula NGC\,3576 (Table 1). The massive cluster is related to the radio H\,II region G\,291.284-0.713,
and its optical counterpart NGC\,3581. The small nebula NGC\,3576 itself has a less massive cluster (Object 65 -- Table 3), thus forming a cluster pair  in that complex.
 Owing to the usual NGC\,3576 designation   in the literature for the massive cluster, we suggest NGC\,3576A for the small one. 
 
 Some of the BRC nebulae (bright-rimmed clouds) have been shown to be sites of star cluster formation (Sugitani et al. 1995).
  The present study indicates 4 BRC clouds with new clusters or stellar groups (Table 3),
   together with 2 BRCs in Table 1.

\section{Discussion}

We compare in Fig. 8 the angular distributions in galactic coordinates  of the present samples (Tables 3 and 4) with those of previously known infrared objects (Table 1) and the optical open clusters (Table 2). The increase of the sample of known clusters and
stellar groups is overwhelming. Note that the infrared clusters coming from the radio nebulae are mainly located between $300^{\circ} < \ell < 350^{\circ}$, corresponding to  internal arms and where absorption in the Galaxy shows a pronounced increase (Dutra \& Bica 2000b).

The distance histograms for the objects obtained from the optical and radio nebula samples (Tables 3 and 4, respectively)
are shown in Fig. 9. For the objects with distance ambiguity we assumed their near distance as a lower limit for the histogram analysis. The histograms show that objects coming from radio nebulae are on the average more distant peaking at $\approx$4.5 kpc, while those from optical
nebulae peak at $\approx$2.5 kpc. We conclude that a typical embedded cluster discovered with 2MASS is not very far
in the Galaxy, but the sampled clusters are located in more internal arms than Sagittarius-Carina, like Scutum-Crux and beyond (e.g. Georgelin \& Georgelin 1976).    

The linear size histograms for the objects obtained from the optical and radio nebula samples (Tables 3 and 4, respectively)
are shown in Fig. 10. The objects with distance ambiguity were excluded. The distributions are similar to that obtained in
Bica et al. (2003). A typical infrared cluster is small (less than 3 pc in linear diameter). The peak for the objects from the radio nebulae is shifted to larger sizes, suggesting that on the average they are more massive clusters. 
At such young ages, star clusters are not dynamically relaxed and their dimensions are not yet modulated by the Galactic tidal field. Consequently, sizes must reflect formation conditions.

\begin{figure} 
\resizebox{\hsize}{!}{\includegraphics{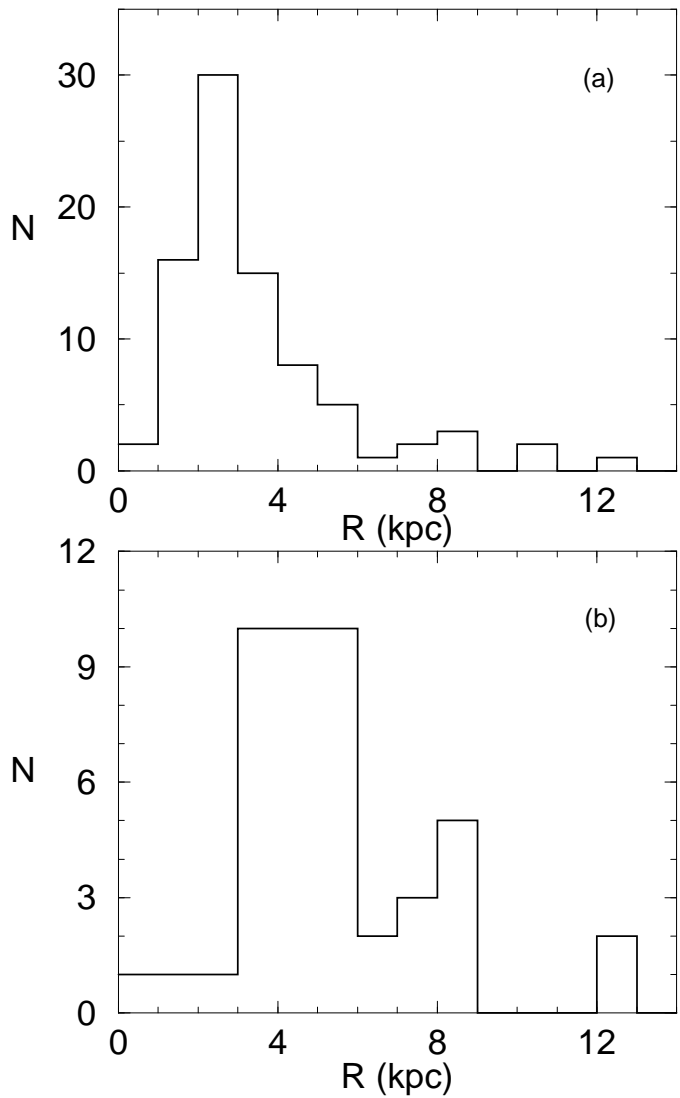}}
\caption[]{Distance histograms for the samples coming from: panel (a)  optical (Table 3) and panel (b) radio nebulae (Table 4).}
\label{fig1}
\end{figure}

An important issue concerns the properties of the optical open clusters 
(Table 2), as compared to those of the IR objects  (Tables 1, 3 and 4), if measured with the same material.
Out of 36 optical open clusters, 17 (Table 2) could be retrieved with the same 
procedures that we used to probe the new IR clusters. 
The sizes for the retrieved optical open clusters measured on  2MASS material with the present
procedures are systematically smaller in the infrared than in the optical (compare Cols. 8 and 11 of Table 2), 
suggesting that we are preferentially seeing denser parts and 
cluster cores in the infrared. 
Probably the optical nebula complex Gum\,55 = RCW\, 113,  owing to the low absorption and proximity,  can clarify the
issue on  stellar component structure in a complex after a few Myr (Table 2).
It contains the populous cluster NGC\,6231 with an optical diameter of $\approx$ 6 pc  and the loose cluster BH\,205 with $\approx$ 2 pc, embedded in  the  extended 
stellar aggregates Harvard\,12 and Tr\,24, with 19 and 14 pc, respectively. 

No prominent cluster was found to be related to radio SNRs. A probable IRGr is in the area of the radio SNR
G330.2+1.0, and one IRCC is in that of G323.5+0.1 (Table 4). Among optical open clusters the association to SNRs is a rare phenomenon too: only Pis\,20 appears to have one (Table 2).

\begin{figure} 
\resizebox{\hsize}{!}{\includegraphics{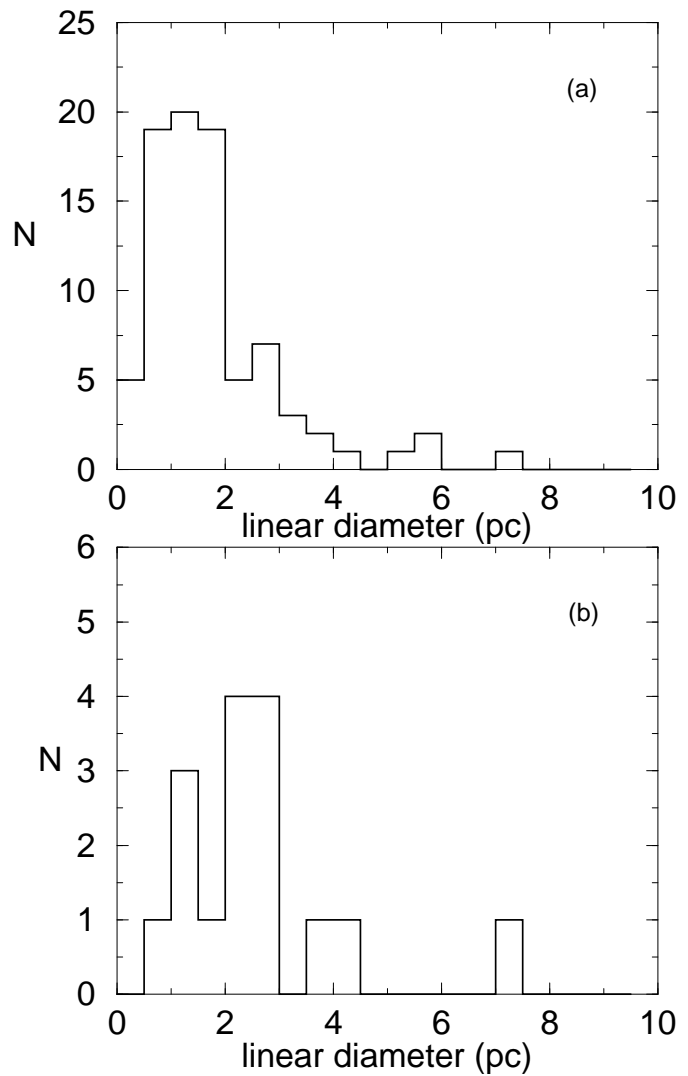}}
\caption[]{Linear major dimension histogram for the samples coming  from: panel (a)  optical (Table 3) and panel (b)  radio nebulae (Table 4).}
\label{fig1}
\end{figure}

Some infrared  clusters and stellar groups occur in pairs or triplets in a fraction of 25\%  
(Bica et al. 2002). For the present sample (Tables 3 and 4) we obtain 21\%, somewhat lower but
suggesting as well a significant role of multiplicy on cluster formation and early dynamical evolution. Taking only
the clusters of Table 3, which are closer on the average (Fig. 9), the resulting fraction is 26\%. For the objects in Table 4 the fraction is 9\%. Small companions would be more difficult to detect in  highly absorbed zones and/or for more distant clusters.

\section{Concluding remarks}

We searched for embedded star clusters 
in the directions of 991 optical and 276 radio nebulae in the Southern Milky Way (encompassing the region
$|b|<10^{\circ}$ , $230^{\circ} <\ell<350^{\circ}$) using 2MASS  all-sky
release JHK$_s$ images. A total of 179 new infrared clusters, stellar groups, candidates and
open clusters were found. They are interesting targets for detailed future
infrared studies. The sample of known infrared clusters and similar objects in the Galaxy is now increased by 
63\%, as compared to  276 objects in  Bica et al.'s (2002) infrared catalogue  together with 10 recent 
additions (Le Duigou \& Kn\"odlseder 2002, Ivanov et al. 2002).

Assuming that the new infrared clusters and stellar groups are physically
linked to the
optical or radio nebulae, we estimated distances and linear
diameters for most of them. On the average objects coming from radio
nebulae are more distant (distribution peak at 4.5 kpc) than those from
optical nebulae (peak at 2.5 kpc). The linear diameter
distributions indicate that objects from the radio nebulae are typically
somewhat  larger than those from optical nebulae, suggesting  that 
they are in general clusters in more massive complexes.

In the studied Southern Milky Way sector we detected  embedded clusters/stellar groups 
related to 12 and 20\% of the catalogued optical
and radio nebulae, respectively.
Considering these detection rates, the remaining nebulae in the disc and
the fact that other
Milky Way sectors have been  more surveyed (Bica et al. 2002), one can
infer
that about 200 infrared star clusters might still be found with the 2MASS
Atlas using the present method. In addition to that search, non-embedded infrared 
clusters should be surveyed by means of
systematic visual inspections and/or automated methods, for a nearly 
complete census of infrared clusters.

\begin{acknowledgements}
This publication makes use of data products from the Two Micron All Sky Survey, which is a joint project of the University of Massachusetts and the Infrared Processing and Analysis Center/California Institute of Technology, funded by the National Aeronautics and Space Administration and the National Science Foundation.
We employed  catalogues from CDS/Simbad (Strasbourg) and  Digitized Sky Survey images from the Space Telescope Science Institute (U.S. Government grant NAG W-2166) obtained using the extraction tool from CADC (Canada). We also made use of the WEBDA open cluster database. We acknowledge support from the Brazilian Institutions CNPq and FAPESP. CMD acknowledges FAPESP for a post-doc fellowship (proc. 00/11864-6).
\end{acknowledgements}

%
%

\end{document}